\documentclass[aps,prb,reprint,superscriptaddress,amsmath,amssymb]{revtex4-2}

\usepackage{graphicx}
\usepackage{bm}
\usepackage{hyperref}
\hypersetup{hidelinks}

\begin{document}

\title{Little-Parks Oscillations and Parity Effects in Mesoscopic Superconducting Rings}

\author{Iogann Tolbatov}
\email[Author to whom correspondence should be addressed: ]{itolbatov@uniss.it}
\affiliation{Department of Chemical, Physical, Mathematical and Natural Sciences, University of Sassari, 07100 Sassari, Italy}

\author{Luca Salasnich}
\affiliation{Dipartimento di Fisica e Astronomia ``Galileo Galilei'' and Padua QTech Center, Universit\`a di Padova, Via Marzolo 8, 35131 Padova, Italy}
\affiliation{INFN Sezione di Padova, Via Marzolo 8, 35131 Padova, Italy}

\begin{abstract}
We analytically investigate the Little-Parks (LP) effect, the periodic oscillation of the superconducting transition temperature $T_c$ driven by an applied magnetic flux $\Phi$, from the weak-coupling Bardeen-Cooper-Schrieffer (BCS) limit to the intermediate-coupling regime in a mesoscopic, quasi-one-dimensional ring. Utilizing a Ginzburg-Landau (GL) expansion with twisted boundary conditions, we correctly incorporate both the discrete finite-size topological parity interference and the continuous center-of-mass (CM) kinetic energy shift induced by the supercurrent. We derive a closed-form expression for $T_c(\Phi)$ valid within the GL framework across variable interaction strengths. We demonstrate that the LP response comprises two distinct physical contributions: a discrete topological parity effect governed by the correlation length ratio $\xi/R$, and a continuous kinetic energy term scaling as $1/R^2$. As the coupling increases and the spatial correlation length of the pairs contracts ($\xi \ll R$), the constituent pairs become geometrically blind to the global boundary constraint, leading to an exponential suppression of the discrete parity modulation. However, the requirement that the condensate acquires a flux-dependent macroscopic CM momentum persists. This ensures that the critical temperature $T_c$ maintains an algebraic LP modulation driven predominantly by the macroscopic kinetic energy of the supercurrent. We bound our theoretical approach by discussing the breakdown of mean-field GL validity before phase fluctuations dominate near the strong-coupling limit.
\end{abstract}

\maketitle

\section{Introduction}

The Little-Parks (LP) effect stands as one of the most striking macroscopic manifestations of quantum mechanics in condensed matter physics. First observed in thin-walled superconducting cylinders in 1962 \cite{LP}, the effect describes the periodic oscillation of the superconducting critical temperature $T_c$ as a function of an applied axial magnetic flux, $\Phi$. Driven by the requirement of single-valuedness of the macroscopic condensate wavefunction, the system must accommodate the magnetic flux by developing a quantized supercurrent. The kinetic energy associated with this macroscopic phase winding quadratically suppresses the condensation energy, leading to a parabolic drop in $T_c$ that periodically resets via topological phase slips at half-integer multiples of the superconducting flux quantum $\Phi_0 = h/2e$. In the mesoscopic regime, this $h/2e$ macroscopic modulation is superimposed over an $h/e$ electronic background. While the Byers-Yang (BY) theorem \cite{BY} permits this $h/e$ periodicity, it is the microscopic finite-size interference that physically drives its non-zero amplitude. The interplay between geometry, boundary conditions, and the LP transition has been extensively explored in mesoscopic metallic structures \cite{Moshchalkov1995, Baelus2001}. Today, the LP effect remains a powerful diagnostic tool for uncovering macroscopic quantum coherence and unconventional pairing mechanisms in modern condensed matter systems, including strongly correlated underdoped cuprates \cite{Liao2022} and materials exhibiting spontaneous $\pi$-phase shifts \cite{Almoalem2024}. Beyond solid-state superconductors, the principles of the LP effect have also been extended to ultracold atomic superfluids \cite{LyandaGeller2000}. Microscopic Gor'kov calculations by Wei and Goldbart \cite{Wei2008} in the weak-coupling BCS limit identified an even-odd parity effect associated with the number of Cooper pairs, yet this discrete phenomenon remained analytically intertwined with the continuous macroscopic $1/R^2$ center-of-mass (CM) kinetic energy. However, extending this analytical separation beyond the asymptotic weak-coupling approximation ($\mu \gg k_B T_c$) to derive the generalized spatial correlation length ($\tilde{\kappa}^{-1}$) across variable coupling strengths remains unexplored.

Over the past two decades, the realization of the BCS to Bose-Einstein condensate (BEC) crossover in ultracold atomic Fermi gases \cite{Regal, Giorgini}, and more recently in tunable two-dimensional solid-state platforms \cite{Nakagawa}, has allowed for the unprecedented tuning of the pairing interaction strength. As the system is tuned from the weak-coupling BCS regime to the strong-coupling BEC regime, the fundamental nature of the condensate transforms. The coherence length $\xi$ shrinks by orders of magnitude.

This evolution raises a question: How does a mesoscopic topological constraint, such as the Aharonov-Bohm (AB) phase winding \cite{AB} of the LP effect, modulate the critical temperature when the constituent pairs become smaller than the geometric radius of the system ($\xi \ll R$)? The interplay between the shrinking coherence length and the geometric curvature of closed topological manifolds governs not only phase boundaries in rings, but also the quantization of collective Higgs and Goldstone excitations in compact superconducting shells \cite{Tolbatov2026Spheres}. As the spatial extent of the pairs shrinks, their sensitivity to the global topological boundary conditions transforms, requiring a treatment of both local quantum interference and macroscopic kinetic energy. In this work, we provide an analytical description of the LP effect from the BCS limit to the intermediate-coupling crossover regime in a mesoscopic ring.
 
Ultimately, we show that as the correlation length shrinks, the discrete finite-size topological parity effect becomes exponentially suppressed. Consequently, the LP modulation of the critical temperature $T_c$ becomes driven almost entirely by the continuous, classical CM kinetic energy of the macroscopic supercurrent, reshaping the phase diagram of tunable mesoscopic superconductors.

\section{Model and Effective Action on a Ring}

We consider an interacting gas of two-component fermions of mass $m$ confined to a quasi-one-dimensional mesoscopic ring of radius $R$ and circumference $L = 2\pi R$. The system is modeled by a continuous contact-interaction Hamiltonian with an attractive $s$-wave pairing interaction $g < 0$ \cite{Yang}. The grand-canonical Hamiltonian reads:

\begin{equation}
\begin{split}
H &= \int_0^L dx \Bigg[ \sum_{\sigma=\uparrow,\downarrow} \psi^\dagger_\sigma(x) \left( -\frac{\hbar^2}{2m}\frac{\partial^2}{\partial x^2} - \mu \right) \psi_\sigma(x) \\
&\quad + g \psi^\dagger_\uparrow(x) \psi^\dagger_\downarrow(x) \psi_\downarrow(x) \psi_\uparrow(x) \Bigg]
\end{split}
\end{equation}
\noindent where $\psi^\dagger_\sigma(x)$ and $\psi_\sigma(x)$ are the standard fermionic creation and annihilation operators for spin $\sigma$ at the longitudinal spatial coordinate $x$, $\hbar$ is the reduced Planck constant, and $\mu$ is the chemical potential. By "quasi-one-dimensional", we imply that the transverse confinement energy heavily exceeds all relevant dynamical scales ($\hbar \omega_\perp \gg \vert\mu\vert, k_B T$, where $\omega_\perp$ is the transverse harmonic confinement frequency, $T$ is the temperature, and $k_B$ is the Boltzmann constant), freezing out transverse kinematics. Consequently, the problem reduces to a 1D Hamiltonian where a true two-body bound state exists for any arbitrarily weak attractive interaction ($g<0$). While the Mermin-Wagner-Hohenberg theorem forbids true long-range order in infinite 1D systems at finite temperatures, in a mesoscopic ring of circumference $L$ the phase correlation length $\xi_\phi(T)$ can exceed $L$ below a characteristic temperature scale. Within the present GL treatment we denote the superconducting instability temperature by $T_c$; fluctuation corrections are not included. Following the standard theoretical framework for ultracold quasi-one-dimensional systems \cite{Guan2013}, the interactions are characterized via the dimensionless interaction parameter $\gamma = m\vert g\vert/(\hbar^2 n)$, where $n = N/L$ represents the 1D density, with $N$ being the total number of fermions. The emergence of Cooper pairs with finite CM momentum in such strongly attractive 1D geometries is a well-established phenomenon, typically treated within the grand canonical ensemble \cite{Guan2013, Liu2007}.

To introduce the LP effect, an external axial magnetic flux $\Phi$ is threaded through the ring. While the electronic system is compatible with the $h/e$ periodicity permitted by the BY theorem, the emergence of a non-zero $h/e$ component in mesoscopic rings is driven by the microscopic finite-size mechanism, consistent with Wei and Goldbart's findings. This finite-size interference ensures that an $h/e$ modulation remains superimposed on the dominant $h/2e$ macroscopic phase-winding of the Cooper pairs. We formally embed this external vector potential into the twisted periodic boundary conditions via the Peierls substitution \cite{Peierls}. For a single fermion of charge $e$, the AB phase enforces:

\begin{equation}
\psi_\sigma(x+L) = e^{i \pi \Phi / \Phi_0} \psi_\sigma(x)
\end{equation}
\noindent where $\Phi_0 = h/2e$ is the superconducting flux quantum.

The mean-field pairing order parameter is defined as $\Delta(x) = g \langle \psi_\downarrow(x) \psi_\uparrow(x) \rangle$. Applying the fermionic boundary conditions to this definition requires the macroscopic order parameter to acquire double the phase twist of the single particles:

\begin{equation}
\Delta(x+L) = e^{i 2\pi \Phi / \Phi_0} \Delta(x)
\end{equation}

To satisfy this topological constraint and minimize the free energy in the presence of flux, the macroscopic superconducting order parameter must develop a persistent supercurrent. The gap function takes the form:

\begin{equation}
\Delta(x) = \Delta_0 e^{i \frac{2\pi}{L}\left(\nu + \frac{\Phi}{\Phi_0}\right) x}
\end{equation}
\noindent where $\nu \in \mathbb{Z}$ is the topological integer winding number. Because the CM momentum operator is $\hat{P} = -i\hbar \partial_x$, acting on this phase gradient implies that the Cooper pairs are globally boosted, possessing a non-zero, flux-dependent macroscopic CM momentum:

\begin{equation}
P_{\mathrm{cm}} = \frac{2\pi\hbar}{L}\left(\nu + \frac{\Phi}{\Phi_0}\right)
\end{equation}

Applying a local $U(1)$ gauge transformation $\tilde{\psi}_\sigma(x) = e^{-i P_{\mathrm{cm}} x / 2\hbar} \psi_\sigma(x)$ maps the order parameter back to a constant real gap ($\tilde{\Delta} = \Delta_0$). However, accommodating this bulk supercurrent physically breaks the symmetry of the Bogoliubov-de Gennes (BdG) spectrum \cite{Gennes}. By applying the phase gradient, the normal-state single-particle kinetic energy becomes $\frac{\hbar^2}{2m}(k + P_{\mathrm{cm}}/2\hbar)^2$. Substituting this expanded form into the BdG matrix yields the eigenvalues:

\begin{equation}
E_{k,\pm} = \frac{\hbar k P_{\mathrm{cm}}}{2m} \pm \sqrt{\left(\xi_k + \frac{P_{\mathrm{cm}}^2}{8m}\right)^2 + \Delta_0^2}
\end{equation}
\noindent where $\xi_k = \frac{\hbar^2 k^2}{2m} - \mu$. Here, the macroscopic uniform supercurrent velocity breaks inversion symmetry via the single-particle Doppler shift $\frac{\hbar k P_{\mathrm{cm}}}{2m}$. The term $\frac{P_{\mathrm{cm}}^2}{8m}$ enters the BdG spectrum as an effective shift to the single-fermion chemical potential ($\mu \to \mu - \frac{P_{\mathrm{cm}}^2}{8m}$). Upon summing over both spin species in the condensed pair, this corresponds to the total macroscopic CM kinetic energy shift $E_{\mathrm{kin, pair}} = \frac{P_{\mathrm{cm}}^2}{4m} = \frac{1}{2}(2m) v_{\mathrm{cm}}^2$ for a composite pair of mass $2m$. By capturing this asymmetric spectrum, the framework accurately evaluates the free energy of a condensate carrying a macroscopic supercurrent velocity, properly establishing the foundational physics of the LP effect.

Integrating out the fermionic degrees of freedom yields the finite-temperature grand potential $\Omega$:

\begin{equation}
\Omega = \frac{L \Delta_0^2}{|g|} + \sum_{k} \left( \xi_k + \frac{P_{\mathrm{cm}}^2}{8m} \right) - \frac{1}{\beta}\sum_{k, \pm} \ln\left(1 + e^{-\beta E_{k,\pm}}\right)
\end{equation}
\noindent where $\beta = 1/(k_B T)$ and the momenta $k$ remain discretized by the ring's finite geometry. Note that the normal-ordering of the BdG matrix also generates a linear odd-momentum cross-term, $\frac{\hbar k P_{\mathrm{cm}}}{2m}$, within the vacuum energy shift. However, because the set of allowed momenta in the ring is symmetric under inversion ($k \to -k$), this odd term cancels in the summation, allowing it to be safely excluded from the thermodynamic trace.

\section{Derivation of the Flux-Dependent $T_c$}

To analytically separate the macroscopic kinetic energy of the supercurrent from the finite-size quantum interference, we expand the grand thermodynamic potential $\Omega(\Delta_0, \Phi, \nu)$ in a Ginzburg-Landau (GL) series \cite{Pujos} near the phase transition ($\Delta_0 \to 0$):

\begin{equation}
\Omega = \Omega_0(\Phi) + a(T, \mu, P_{\mathrm{cm}}, \Phi) \Delta_0^2 + \frac{b}{2} \Delta_0^4
\end{equation}
where the equality holds to leading order in the order parameter $\Delta_0$ near the phase transition.

Here, $T_c(\Phi, \nu)$ denotes the flux-dependent critical transition temperature evaluated within the mean-field GL framework at the boundary $a(T_c, \mu, P_{\mathrm{cm}}, \Phi) = 0$. It is important to note that this derivation is performed in the grand-canonical ensemble, with the chemical potential $\mu$ treated as an external control parameter. This approach cleanly isolates the thermodynamic shift induced by the macroscopic kinetic energy of the supercurrent from the discrete topological interference, avoiding the need to simultaneously resolve the number equation $n = -\frac{1}{L}\frac{\partial\Omega}{\partial\mu}$. By expanding the gap equation and carefully separating the Doppler shift $P_{\mathrm{cm}}$ from the discrete density of states, the GL quadratic coefficient takes the form:

\begin{equation}
\begin{split}
a(T, \mu, P_{\mathrm{cm}}, \Phi) &= a_{\mathrm{bulk}}(T, \mu) + a_{\mathrm{kin}}(T, \mu) P_{\mathrm{cm}}^2 \\
&\quad + \delta a_{\mathrm{FS}}(T, \mu, \nu)
\end{split}
\end{equation}

Here, $a_{\mathrm{bulk}}$ defines the zero-flux bulk critical transition temperature $T_{c,0}$. The remaining two terms govern the LP response. The macroscopic kinetic energy term, $a_{\mathrm{kin}}$, is obtained by expanding the pair susceptibility $\chi$ to second order in $P_{\mathrm{cm}}$. This captures the standard macroscopic GL kinetic energy of the supercurrent, which scales with the superfluid density, a response dictating the phase stiffness and the energetic cost of carrying a finite CM momentum across the crossover \cite{Tolbatov2026Altermagnetism}, imposing a purely continuous algebraic suppression:

\begin{equation}
a_{\mathrm{kin}} P_{\mathrm{cm}}^2 = \frac{\mathcal{C}}{R^2}\left(\nu + \frac{\Phi}{\Phi_0}\right)^2
\end{equation}
\noindent where $\mathcal{C}$ is a temperature- and chemical-potential-dependent prefactor corresponding to the superfluid phase stiffness.

The second contribution, the finite-size parity effect $\delta a_{\mathrm{FS}}$, captures the discrete topological interference on the single-particle wavefunctions. To evaluate this, we must track the allowed momentum states. Following the gauge transformation $\tilde{\psi}_\sigma(x) = e^{-i P_{\mathrm{cm}} x / 2\hbar} \psi_\sigma(x)$, the single-particle boundary condition becomes:

\begin{equation}
\tilde{\psi}_\sigma(x+L) = e^{-i P_{\mathrm{cm}} (x+L) / 2\hbar} \psi_\sigma(x+L)
\end{equation}

Substituting the original twisted boundary condition $\psi_\sigma(x+L) = e^{i \pi \Phi / \Phi_0} \psi_\sigma(x)$ alongside the macroscopic momentum $P_{\mathrm{cm}} L / 2\hbar = \pi(\nu + \Phi/\Phi_0)$ yields:

\begin{equation}
\tilde{\psi}_\sigma(x+L) = e^{-i \pi (\nu + \Phi/\Phi_0)} e^{i \pi \Phi / \Phi_0} \tilde{\psi}_\sigma(x) = e^{-i \pi \nu} \tilde{\psi}_\sigma(x)
\end{equation}

Since the topological winding number $\nu$ is an integer, this simplifies to a purely discrete parity constraint:

\begin{equation}
\tilde{\psi}_\sigma(x+L) = (-1)^\nu \tilde{\psi}_\sigma(x)
\end{equation}

This cancellation of $\Phi$ in the boundary condition of the gauge-transformed field $\tilde{\psi}_\sigma$ serves as a proof that there is no double-counting of the magnetic flux. The continuous flux $\Phi$ is entirely absorbed by the macroscopic Doppler shift in the BdG Hamiltonian (yielding $a_{\mathrm{kin}}$), leaving the boundary condition to depend solely on the discrete topological winding integer $\nu$ (yielding $\delta a_{\mathrm{FS}}$). Consequently, gauge invariance is preserved, and the continuous AB flux $\Phi$ is stripped from the relative internal momentum of the Cooper pairs. The allowed quantized momenta for the transformed field are $k_m = \frac{2\pi}{L}\left(m + \frac{\nu}{2}\right)$, independent of $\Phi$. Applying the Poisson summation formula \cite{Cheung} with the momentum index shift $c = \frac{\nu}{2}$ extracts the finite-size corrections $\delta a_{\mathrm{FS}}$ as a discrete parity effect governed by Fourier harmonics $e^{i 2\pi w (\nu/2)}$, which simplify to $(-1)^{w\nu}$:

Retaining only the leading-order terms at $P_{\mathrm{cm}}=0$, the finite-size corrections $\delta a_{\mathrm{FS}}$ evaluate to:
\begin{equation}
\begin{split}
\delta a_{\mathrm{FS}}(T, \nu) &= \lambda_0 \sum_{w=1}^{\infty} (-1)^{w\nu} \\
&\quad \times \int_{-\infty}^{\infty} dk \, \chi\left(\frac{\hbar^2 k^2}{2m}, \mu, T\right) \cos(w k L)
\end{split}
\end{equation}
\noindent where $\lambda_0$ is an effective coupling constant. To control the separation of kinetic and parity effects, the Poisson summation must be applied to the complete susceptibility $\chi(k, P_{\mathrm{cm}})$. Expanding the finite-size susceptibility in powers of the CM momentum $P_{\mathrm{cm}}$ yields:
\begin{equation}
\delta a_{\mathrm{FS}} = \sum_{w=1}^{\infty} (-1)^{w\nu} \left[ \mathcal{I}_w^{(0)} + P_{\mathrm{cm}}^2 \mathcal{I}_w^{(2)} + \mathcal{O}(P_{\mathrm{cm}}^4) \right]
\end{equation}
While $\mathcal{I}_w^{(0)}$ dictates the discrete parity effect at zero flux, the mixed term $P_{\mathrm{cm}}^2 \mathcal{I}_w^{(2)}$ cannot be neglected trivially. Because $P_{\mathrm{cm}}$ is inherently tied to the continuous magnetic flux $\Phi$, the cross-term $P_{\mathrm{cm}}^2 \mathcal{I}_w^{(2)}$ is precisely the mechanism that preserves the continuous flux-dependence of the finite-size correction.

While Wei and Goldbart \cite{Wei2008} correctly identified an even-odd parity effect for Cooper pairs, their theoretical framework treats this phenomenon as an amplitude modulation of the continuous flux. As shown in their findings (e.g., their Eq. 41), the finite-size parity factor is inextricably coupled to the continuous flux via a trigonometric multiplier, physically corresponding to our mixed cross-term $P_{\mathrm{cm}}^2 \mathcal{I}_w^{(2)}$. By analytically tracking both the pure parity ($\mathcal{I}_w^{(0)}$) and continuous cross-terms ($\mathcal{I}_w^{(2)}$) away from the asymptotic $\mu \gg k_B T_c$ limit, our formulation reveals that the mixed term is parametrically suppressed by the macroscopic scaling factor $\hbar^2 / (2m R^2 k_B T_c) \ll 1$. Because the ring circumference bounds the macroscopic kinetic energy to be vastly subordinate to the thermal coherence scale in the mesoscopic regime, $P_{\mathrm{cm}}^2 \mathcal{I}_w^{(2)}$ scales out of the leading-order free energy. Consequently, the Cooper pair parity effect is not merely a flux-dependent envelope (as necessitated in prior limits), but uniquely decouples into a pure, standalone discrete topological shift.

This demonstrates that in a one-dimensional ring, the continuous parabolic LP modulation is governed by the macroscopic kinetic energy $a_{\mathrm{kin}} P_{\mathrm{cm}}^2$. The dominant finite-size interference contributes as a discrete parity effect, shifting the baseline $T_c$ depending on whether the winding number $\nu$ is even or odd. Because advancing the flux by $\Phi_0 = h/2e$ shifts the winding number by $\Delta\nu = 1$, the parity effect flips between even and odd states. Therefore, the parity state of the system only resets when the flux advances by $h/e$ ($\Delta\nu = 2$). This ensures the system remains compatible with the $h/e$ periodicity permitted by the BY theorem, upon which the emergent $h/2e$ macroscopic modulation of the condensate is superimposed.

Setting the total coefficient to zero yields the analytical expression for the flux-dependent critical temperature valid within the perturbative limit of the GL expansion:

\begin{equation}
T_c(\Phi, \nu) = T_{c,0} - \frac{a_{\mathrm{kin}}(T_{c,0}) P_{\mathrm{cm}}^2 + \delta a_{\mathrm{FS}}(T_{c,0}, \nu)}{a'_{\mathrm{bulk}}(T_{c,0})}
\label{eq:Tc}
\end{equation}

The physical LP envelope observed in experiment is the maximum $T_c$ over all integer winding numbers $\nu$. The oscillation is driven by both the macroscopic $1/R^2$ continuous kinetic term and the discrete finite-size parity shift.

\section{Asymptotic Limits and Crossover Behavior}

\begin{figure*}[t]
    \centering
    \includegraphics[width=0.85\textwidth]{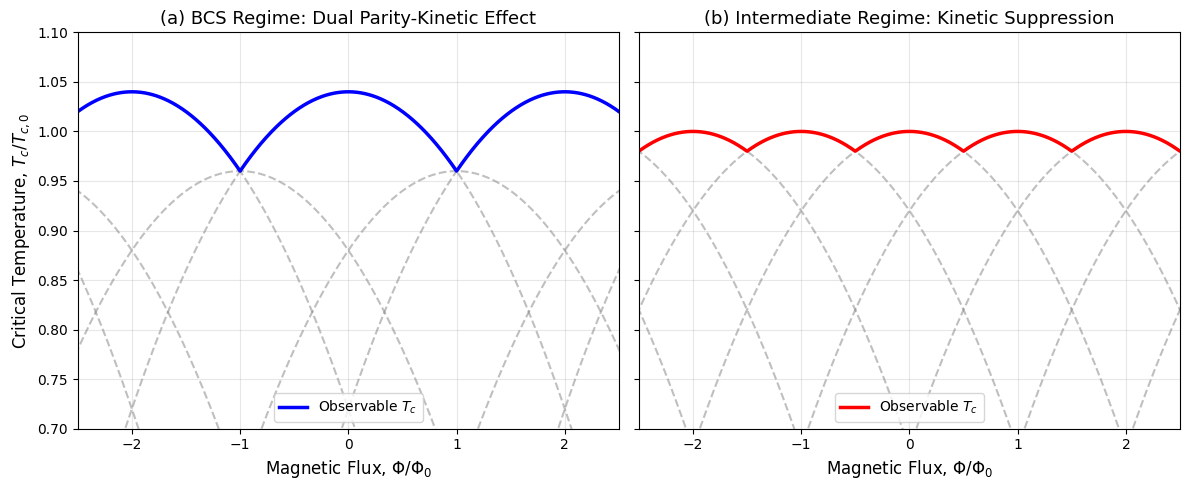}
    \caption{Phenomenological illustration of the LP critical temperature ($T_c$) envelope (thick solid lines) as a function of magnetic flux, derived from Eq. (\ref{eq:Tc}). Note that this plot serves as a phenomenological illustration rather than a quantitative universal trend. The dashed gray lines represent the underlying available energy states for different integer winding numbers $\nu$. \textbf{(a)} In the weakly coupled BCS regime, the $T_c$ modulation exhibits the dual parity-kinetic effect: the macroscopic kinetic energy of the supercurrent creates the parabolic envelope, while the finite-size AB interference introduces a discrete baseline shift between even and odd $\nu$ states. \textbf{(b)} In the intermediate-coupling regime, the spatial contraction of the susceptibility correlation length exponentially suppresses the finite-size parity effect. The surviving LP modulation of the critical temperature $T_c$ is predominantly algebraic, driven by the $P_{\mathrm{cm}}^2/4m$ CM kinetic energy of the macroscopic winding.}
    \label{fig:LP_crossover}
\end{figure*}

The evolution of the LP amplitude from the BCS to the intermediate-coupling regime is governed by the behavior of both the macroscopic kinetic term and the finite-size spatial interference integral:
\begin{equation}
\mathcal{I}_w = \int_{-\infty}^{\infty} dk \, \chi(k, \mu, T_c) \cos(w k L)
\end{equation}
where $\chi(k, \mu, T_c)$ represents the static pair-channel response kernel evaluated at $T = T_c$:
\begin{equation}
\begin{split}
\chi(k, \mu, T_c) &= \frac{1}{\beta}\sum_{\omega_n} \frac{1}{\left(i\hbar\omega_n - \xi_k\right)\left(-i\hbar\omega_n - \xi_k\right)} \\
&= \frac{\tanh\left(\frac{\xi_k}{2 k_B T_c}\right)}{2 \xi_k}
\end{split}
\end{equation}
with the single-particle dispersion $\xi_k = \frac{\hbar^2 k^2}{2m} - \mu$.

In the weak-coupling BCS limit ($\mu = E_F \gg k_B T_c$), the extended correlation length ($\xi \gg R$) ensures that both LP terms are active. The finite-size integral evaluates via contour integration to $\mathcal{I}_w \propto \exp(-w L/\xi)$, with the exponential factor approaching unity in the asymptotic limit $L/\xi \to 0$. Consequently, the transition temperature is highly susceptible to the topological winding, leading to sharp, intersecting parabolas dominated continuously by the macroscopic $1/R^2$ kinetic energy, with the finite-size interference imposing an additional discrete even/odd parity shift on the amplitude.

While previous weak-coupling BCS treatments of superconducting rings, notably by Wei and Goldbart \cite{Wei2008}, successfully identified an exponentially suppressed $h/e$ finite-size correction, their analytical results rely on the asymptotic approximation $\mu \gg k_B T_c$. Here, our objective is to retain the full quadratic dispersion and evaluate the pair susceptibility without this approximation.

By utilizing the pair susceptibility $\chi(k, \mu, T_c) = \frac{\tanh(\xi_k / 2k_B T_c)}{2\xi_k}$, the leading behavior is governed by its principal complex poles in the upper half of the complex plane, $k = \pm q + i\tilde{\kappa}$. For $\mu > 0$, these evaluate to:
\begin{align}
q &= \sqrt{\frac{m}{\hbar^2}\left(\sqrt{\mu^2 + (\pi k_B T_c)^2} + \mu\right)}, \\
\tilde{\kappa} &= \sqrt{\frac{m}{\hbar^2}\left(\sqrt{\mu^2 + (\pi k_B T_c)^2} - \mu\right)}.
\end{align}
Consequently, the finite-size contribution yields the analytical form $\mathcal{I}_w = A(\mu, T_c) e^{-\tilde{\kappa} w L} \cos(q w L - \varphi)$, where this equality strictly captures the leading asymptotic behavior dictated by the principal poles. Beyond the asymptotic weak-coupling limit, our treatment provides the generalized expressions for both the decay rate $\tilde{\kappa}$ and the oscillation wavenumber $q$. The resulting finite-size correction retains an oscillatory dependence and can change sign as a function of $qL$.

To establish analytical continuity with the results of Wei and Goldbart \cite{Wei2008}, we take the BCS limit $\mu \gg k_B T_c$. Expanding our expression for $\tilde{\kappa}$, we obtain $\tilde{\kappa} = \frac{\pi k_B T_c}{\hbar v_F}$, where the equality holds to leading order in $k_B T_c / \mu$. Using standard BCS conventions, the zero-temperature coherence length is defined as $\xi_0 = \frac{\hbar v_F}{\pi \Delta_0}$, where the zero-temperature gap relates to $T_c$ via $\Delta_0 = \Gamma k_B T_c$ with $\Gamma = \pi e^{-\gamma} = 1.764$.

Substituting this yields $\tilde{\kappa} = \frac{1}{\Gamma \xi_0}$. The leading finite-size exponential penalty $e^{-\tilde{\kappa} L}$ therefore perfectly reduces to $\exp\left(-\frac{2\pi R}{\Gamma \xi_0}\right)$ for $L=2\pi R$. This analytical matching demonstrates how the weak-coupling exponential scale of Wei and Goldbart is recovered as the asymptotic limit of our generalized pole formulation.

As the interaction strengthens and $\mu$ contracts toward the crossover ($\mu \to 0$), $\tilde{\kappa} \to \sqrt{m \pi k_B T_c / \hbar^2}$. The correlation length contracts such that $\tilde{\kappa} R \gg 1$. Consequently, the exponential factor $e^{-w \tilde{\kappa} L}$ heavily suppresses the finite-size parity correction. The constituent pairs become locally insensitive to the global topological boundary conditions of the ring. However, the CM kinetic term $a_{\mathrm{kin}} P_{\mathrm{cm}}^2$ survives unaltered, as the macroscopic order parameter must still carry a global phase gradient to accommodate the threaded flux.

We emphasize that extending this mean-field derivation of $T_c$ into the deep strong-coupling regime is precluded in quasi-1D geometries. In that limit, strong thermal phase fluctuations dominate \cite{Langer1967, McCumber1970}, decoupling $T_c$ from the mean-field pairing scale and rendering a GL expansion ill-posed.

\section{Experimental Feasibility}

The transition from a dual parity-kinetic LP effect (BCS) to an increasingly algebraic kinetic suppression provides a distinct, experimentally observable signature of the shrinking coherence length. In ultracold atomic platforms, as the system is tuned toward intermediate coupling, experiments could observe the exponential suppression of the discrete finite-size parity ripples, leaving behind a smooth, continuous algebraic $1/R^2$ modulation of the pairing instability onset $T_c$. While our calculated $T_c$ identifies the mean-field pairing instability, we note that in physical quasi-1D systems, the observable zero-resistance transition may be further broadened by thermal fluctuation effects not captured here.

Furthermore, this paradigm can also be explored in lithographically defined mesoscopic rings of moiré superlattices, such as magic-angle twisted bilayer graphene (MATBG) \cite{Cao}, where the effective pairing strength (parameterized by the dimensionless coupling ratio $\Delta/E_F$) is continuously tuned via electrostatic gating of the carrier density relative to the flat-band singularity. By utilizing precision magnetotransport to track the global zero-resistance transition, experimentalists can observe the evolution of the LP amplitude. As the device is electrostatically tuned toward the stronger-coupling flat-band regime, transport measurements should reveal the dampening of the discrete parity ripples, with the macroscopic transition continuing to oscillate predominantly via the algebraically suppressed CM kinetic energy $P_{\mathrm{cm}}^2/4m$ of the supercurrent. We note, however, that observing the discrete parity effect in solid-state platforms requires the ring's circumference to remain smaller than the elastic mean free path ($L < l_e$). In lithographically etched devices, edge-induced disorder and intervalley scattering can rapidly suppress the ballistic single-particle quantum interference required for $\delta a_{\mathrm{FS}}$, potentially masking the parity effect via Dingle-like suppression before the intermediate-coupling regime is reached.

\section{Conclusion}

In summary, we have presented an analytical theory of the LP effect from the BCS limit to the intermediate-coupling crossover regime in a mesoscopic superconducting ring. By resolving the correct topological boundary conditions for the macroscopic pairing order parameter, we explicitly incorporated the pair-breaking Doppler shift required to support a supercurrent. This allowed us to analytically separate the finite-size AB quantum interference from the macroscopic kinetic energy of the condensate.

Our central finding is that the character of the LP oscillation profoundly transforms as the system evolves from extended Cooper pairs toward tighter spatial correlations. Consistent with the weak-coupling BCS findings of Wei and Goldbart, the continuous parabolic suppression of $T_c$ is driven by the macroscopic kinetic energy. We demonstrated that because the continuous flux-dependent finite-size cross terms are parametrically suppressed by the macroscopic scale of the ring ($\hbar^2 / 2mR^2 \ll k_B T_c$), they are strictly higher-order. This physical scale separation mathematically justifies decoupling the finite-size interference into a pure, leading-order discrete topological parity effect, independent of the continuous flux modulation. However, by extending this separation to intermediate coupling without assuming $\mu \gg k_B T_c$, we have shown that the spatial contraction of the pair susceptibility correlation length causes this discrete finite-size parity effect to become exponentially suppressed. Despite this local geometric blindness, the macroscopic LP effect does not disappear. The topological burden is absorbed by the CM coordinate of the pairs, maintaining a continuous $1/R^2$ LP modulation of the critical temperature $T_c$. A full quantitative description of the deep BEC limit, where intense thermal phase fluctuations ultimately invalidate the mean-field paradigm in low-dimensional systems, will require theoretical frameworks that move beyond local actions to properly capture vortex confinement and long-range phase coherence \cite{Tolbatov2026Superdiffusive}. Such extensions remain a compelling direction for future theoretical studies.

\begin{acknowledgments}
L.S. is partially supported by the Project ``Frontiere Quantistiche'' (Dipartimenti di Eccellenza) of the Italian Ministry of University and Research and by ``Iniziativa Specifica Quantum'' of INFN.
\end{acknowledgments}

\end{document}